\documentclass[conference,compsoc]{IEEEtran}
\ifCLASSOPTIONcompsoc
  \usepackage[nocompress]{cite}
\else
  \usepackage{cite}
\fi
\ifCLASSINFOpdf
\else
\fi
\usepackage{dirtytalk}
\usepackage{hyperref}
\usepackage{graphicx}

\IEEEoverridecommandlockouts 
\begin{document}
\title{Summarizing and Analyzing the Privacy-Preserving Techniques in Bitcoin and other Cryptocurrencies}

\author{\IEEEauthorblockN{Chaitanya Rahalkar}
\thanks{Both the authors have contributed equally}
\IEEEauthorblockA{School of Computer Science\\
Georgia Institute of Technology\\
cr@gatech.edu}
\and
\IEEEauthorblockN{Anushka Virgaonkar}
\IEEEauthorblockA{School of Computer Science\\
Georgia Institute of Technology\\
avirgaonkar3@gatech.edu}
}


% compsoc conferences compared to traditional conferences), use this
%Telephone: (800) 555--1212, Fax: (888) 555--1212}


\maketitle

\begin{abstract}
Bitcoin and many other similar Cryptocurrencies have been in existence for over a decade, prominently focusing on decentralized, pseudo-anonymous ledger-based transactions. Many protocol improvements and changes have resulted in new variants of Cryptocurrencies that are known for their peculiar characteristics. For instance, Storjcoin is a Proof-of-Storage-based Cryptocurrency that incentivizes its peers based on the amount of storage owned by them \cite{storj2018storj}. Cryptocurrencies like Monero strive for user privacy by using privacy-centric cryptographic algorithms \cite{van2013cryptonote}. 
While Cryptocurrencies strive to maintain peer transparency by making the transactions and the entire ledger public, user privacy is compromised at times. Monero and many other privacy-centric Cryptocurrencies have significantly improved from the original Bitcoin protocol after several problems were found in the protocol. Most of these deficiencies were related to the privacy of users. Even though Bitcoin claims to have pseudo-anonymous user identities, many attacks have managed to successfully de-anonymize users. In this paper, we present some well-known attacks and analysis techniques that have compromised the privacy of Bitcoin and many other similar Cryptocurrencies. We also analyze and study different privacy-preserving algorithms and the problems these algorithms manage to solve. Lastly, we touch upon the ethics, impact, legality, and acceptance of imposing these privacy algorithms. 
\end{abstract}

\IEEEpeerreviewmaketitle

\section{Introduction}
Since its inception in 2009, Bitcoin has been heavily studied by researchers to look for flaws and improvements. Undoubtedly, Bitcoin has been one of the most successful Cryptocurrencies. Consequently, it has been targeted by adversaries and is under constant surveillance by government entities. Many attacks have been attempted on different aspects of the protocol like transaction malleability, double-spending, block withholding etc. However, in the privacy domain, multiple issues have been found in the protocol that attacks the pseudo-anonymity claims of Bitcoin. Several hard-forks or completely new Cryptocurrencies have been introduced that target these privacy problems in various ways. Bitcoin uses 58 character (Base58 encoded) addresses which are cryptographically linked to a user's public-private key pair. An address is essentially a hash of the owner's public key. A public-private key-pair can be used by a user (or peer) to claim authority over his / her Bitcoins. Bitcoin uses these addresses to perform transactions on the blockchain. Many attacks have been attempted that have managed to trace back addresses to their owners. Also, government surveillance agencies constantly monitor the chain for high-value transactions and attempt to trace back the origin of the transaction. Cryptocurrency exchanges are the exit points where the Cryptocurrency ownership can be found out. Many exchanges require their users to verify themselves through IDs (like SSN) before Bitcoins are converted to a fiat currency.  T. Okamoto and K. Ohta, in their paper on \say{Universal Electronic Cash} specified two important properties that a Cryptocurrency model must adhere to - untraceability, and unlikability \cite{okamoto1991universal}. Untraceability means that for every incoming transaction, all possible senders are equiprobable. Unlinkability implies, for any two outgoing transactions, it is impossible to prove they were sent to the same person. Unfortunately, Bitcoin does not adhere completely to these two properties. For the scope of this paper, we can consider that any Cryptocurrency that manages to satisfy these two properties can be said to have achieved complete anonymity. Although there have been attempted attacks on privacy-respecting currencies like Monero, they are beyond the scope of this research. 
\section{Tiers of Privacy}
Based on the two identified properties of untraceability and unlinkability, anonymity can be classified into four tiers \cite{SOK}. Based upon how effectively a Cryptocurrency manages to satisfy the two stated properties there can be four classes - 
\begin{enumerate}
    \item Pseudonymity - This kind of anonymity is achieved through pseudonymous addresses which are typically used in Bitcoin. It is a disguised state, intermediate between full anonymity and open information. 
    \item Set anonymity - In set anonymity, the identity of a user is either $1$ out of $n$ possible peer identities. Set anonymity is used in the Monero Cryptocurrency through the usage of Ring Signatures proposed by Ron Rivest, Adi Shamir, and Yael Tauman Kala at ASIACRYPT \cite{bender2006ring}. Here $n$ is the ring size. Ring signatures are explained elaborately in the later sections of this paper. Cryptocurrency tumblers or mixers also use set anonymity. 
    \item Full anonymity - This tier of anonymity provides complete concealment of the sender node, the receiver node, and the details of the transaction. As an example, Zerocoin protocol was suggested as a proposed extension to the Bitcoin protocol that strives to provide complete anonymity. It proposed a mechanism to perform transactions by using Zero-knowledge proofs. Transactions that use Zerocoin are drawn from an escrow pool \cite{miers2013zerocoin}. The complete transaction history of the coin is erased when it emerges. Zerocash, a protocol that was an extension of Zerocoin, further improved privacy by concealing the amount that was transacted. 
    \item Confidential Transactions - This tier of anonymity focuses on obfuscating the transaction amount to prevent analysis or any kind of inference attacks. Monero offers transaction amount privacy through the CryptoNote protocol. Even though this cannot be called a tier, many Cryptocurrencies have attempted to exclusively implement confidential transactions and disregard other tiers of anonymity (full anonymity, etc.)
\end{enumerate}
\section{Privacy Attacks on Bitcoin}
Bitcoin falls in the \say{Pseudonymity} tier from the four distinctly defined tiers of privacy. Many privacy attacks have been attempted on the Bitcoin blockchain by finding loopholes or exploiting the evident facts/limitations of the protocol. The attacks listed in this section are not limited to the Bitcoin blockchain but can be attempted on similar kinds of blockchains (e.g. Ethereum) or hard-forks. The privacy attacks attempt to violate either one or both properties (untraceability and unlinkability) of a privacy-preserving transaction system.   
\subsection{Traceability with Transaction Graphs}
Among the different types of privacy attacks attempted on the Bitcoin blockchain, the most prominently known is transaction traceability using transaction graphs. Many tools exist on the internet that create linkages between specified addresses and their corresponding transactions using the transaction graph and the publicly available ledger. If the pseudo-anonymous owner of any one of the addresses in the graph is known, transactions can be traced back. Most often, peers use more than one address to transact on the network. The previous transaction history of these addresses constructed with this graph can help link these addresses to their owner. Several clustering techniques have been used as well, to de-anonymize the peers. However, this attack becomes more difficult if a new address is used for every transaction. 

\subsection{Common-input-ownership Heuristic}
% Coinjoin in BTC - no changes to the protocol
Bitcoin is based on the UTXO model, which maintains a set of unspent transactions. These unspent transactions are sources of inputs to future transactions. Typically a Bitcoin transaction has multiple inputs and outputs. A common heuristic or assumption can be made that if a transaction has more than one input, these can be typically owned by one entity. Even though there exists a possibility of multiple users performing the transaction (multi-sig transaction by multiple users), the chance of it being by the same owner cannot be neglected. Since Bitcoin wallets are known to manage hot-cold addresses, an owner is likely to transact by collectively using amounts from multiple owned addresses. 
\subsection{Wallet Fingerprinting}
Typically, several techniques are used by wallet software to create a unique fingerprint while making transactions. These fingerprinting techniques are listed as follows - 
\begin{enumerate}
    \item Coin Selection - Various wallet software use predefined algorithms to decide which available UTXOs to spend. These distinctly defined algorithms can be used to fingerprint the software wallets. 
    \item Key Storage - Many older wallets do not use compression to store the public keys. However, the newer wallet software use some kind of compression algorithm to store the public keys. This knowledge can be used to distinctly fingerprint the kind of wallet software used. 
    \item Inclusion of nLockTime in the Transactions - Some wallet software include the nLockTime parameter in the transaction fields to ensure that the transaction is locked and not included into the blockchain until the lock time parameter expires. This is typically used to prevent fee sniping attacks. 
    \item Address Formats - Most wallet software use a single address type to perform transactions. Most of the time, this address format is of the Pay-to-public-key hash. Therefore, if any transaction includes a different kind of address format (like Pay-to-script hash) it is likely to be a change address or of a different kind of wallet. 
\end{enumerate}
The reasons to create these unique fingerprints may or may not be intentional. These fingerprints can reveal significant information about the wallet software that is being used as these techniques can be proved to distinctly identify the wallet type. Careful analysis of some of these fingerprinting parameters can also lead to some useful inferences about the sender and the receiver or any intermediaries that are involved. 

\subsection{Round Numbers}
Transaction amounts like 1.5BTC, 0.5BTC are round-number transactions. At times, a transaction that is made using a non-round number amount can be a round number in a different Cryptocurrency or a fiat currency. As of writing this paper, 0.000016 BTC is equivalent to 1 USD. So, if a transaction of 0.000016 BTC is made, it is likely that the transaction was made to a Bitcoin exchange to convert the Bitcoin to a fiat currency. (US dollars in this case). It can also help in knowing change addresses which most likely may belong to the same user. Typically, by knowing Cryptocurrency-fiat conversion rates, the destination country or region can also be analyzed. 

\subsection{Taint Analysis}
Pseudo-anonymity in Bitcoin can be exploited with taint analysis of transactions in Bitcoin. If the address of a user is known to an adversary (e.g. the address might be posted online on the user's personal website) and the outflow of transactions is tracked from this user, then the end-users to which the transactions are being made, can be said to be \say{taintted} with coins from this user. This scheme can be particularly useful when the origin of coins for a particular user is not clearly known. Taint analysis can help in understanding the flow of the coins. This analysis can be affected by performing random transactions with arbitrary amounts to generate noise in the network. 

\subsection{Dust Attacks}

Dust transactions are the transactions resulting from the transfer of an amount that is so small that it is impractical to cover the transaction fee for processing the transaction\cite{DustTransactions}. It can cause unnecessary and avoidable delays in proposing blocks in the blockchain. They take up space on the blockchain that can and should have been used for legitimate transactions. Attackers might use a large number of dust transactions to perform DoS-like attacks on the network, causing congestion and delays in the network, preventing legitimate users from being able to mine blocks. Dust transactions can be leveraged by an adversary to attack the privacy of a user. This is done by the execution of a dust attack. An adversary transfers dust to multiple addresses on the blockchain. An unsuspecting user, who gets sent the dust amount on multiple addresses, controlled by the same user, at some point in time would sweep the dust amount and collect it in one wallet. The adversary would conduct taint analysis on the dust and can determine the owner of a set of public addresses. The adversary with or without obtaining additional information from the blockchain and the real-world can build user-profiles and may succeed in de-anonymizing the user\cite{CryptoDust}.\\

For several reasons, it may seem that there is no reason for a legitimate sender to send dust transactions. However, there are a few legitimate and meaningful use cases for dust transactions:
\begin{enumerate}
    \item Law enforcement agencies use dusting attacks to track down individuals that are involved in illegal activities such as money laundering, criminal activities, etc.
    \item Some organizations like DustAid, ask users to collect dust present in their accounts and donate it to charity.
    \item Dusting can be useful for testing the blockchain network or for academic research purposes.
\end{enumerate}

\subsection{Equal Output CoinJoin Transactions}
Equal Output CoinJoin transactions are known to reveal the change address when the outputs are not equal valued. 
Consider an example - 
\begin{verbatim}
              A (1BTC)
X (6BTC) ---> B (3BTC)
Y (2BTC)      C (3BTC)
              D (1BTC)
\end{verbatim}
In this example, outputs B and C clearly correspond to input X and outputs A and D correspond to input Y. Therefore, either B or C can be the change addresses for input X and A or D can be change addresses for Y. Bitcoin fixes this problem of common-input-output heuristics through CoinJoin. 

\section{Countermeasures Used in Bitcoin}
Bitcoin does provide some in-place countermeasures that can tackle the problem of privacy. Some of these features have been incorporated within the blockchain through soft forks and some of them are still on the soft-fork wishlist. Not all of the measures presented in this section have been incorporated or actively used due to some known problems with the proposals. However, these techniques have proved to be significantly impactful from the privacy perspective, when we disregard other factors like performance, large storage capacity on the chain, etc. Even though some of these techniques are implemented, not all entities involved in the transactions incorporate these measures. These entities can be considered as \say{weak-links} and can allow attackers or surveillance agencies to perform inferences. Collectively incorporating these countermeasures can tackle many of the known privacy problems with Bitcoin. Some of the prominently known countermeasures are listed in this section. 
\subsection{CoinJoin}
\begin{figure}[hbt!]
\centering
\includegraphics[width=2.5in]{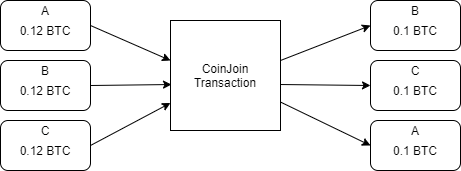}
\caption{CoinJoin Transaction}
\label{coinjoin}
\end{figure}

CoinJoin is one of the techniques that significantly improve the privacy of the transaction by combining inputs from multiple senders into one single transaction. This makes it considerably difficult for any non-involved entity to trace back the transactions and make inferences. It can be effectively used to break the common-input-output heuristic problem listed in 2.2. CoinJoin involves a trusted coordinator (when a centralized server is used) that performs the transaction. It offers privacy through deniability. Looking at the transaction, one can tell whether a CoinJoin transaction has been performed or not. However, it can't be told which participant owns which output. Essentially, CoinJoin allows us to destroy UTXOs and create new ones. The only link between the old and the newly created UTXOs is the CoinJoin transaction. Figure \ref{coinjoin} shows three inputs to the CoinJoin transaction. Here the three inputs are 0.12 BTC of the three participants and the three outputs are 0.1 BTC (here 0.02 BTC is the transaction fee given to the miner). The idea of CoinJoin was proposed by Greg Maxwell % https://bitcointalk.org/?topic=139581 % 
It can be explained in the following steps -
\begin{enumerate}
    \item Find peers interested in participating in CoinJoin.
    \item Exchange the input and output addresses.
    \item Construct the transaction. (Done by the trusted coordinator).
    \item Every participant signs their respective input and checks if their output is present in the transaction or not. 
    \item Broadcast the transaction to the other nodes to get it validated. 
\end{enumerate}

As promising as CoinJoin looks, it comes with several other privacy issues. CoinJoin does expect its participants to carry out the transactions over an anonymous network like Tor, I2P, etc. since the transaction can likely result in leakage of the IP addresses of its participants. CoinJoin suffers from a major problem of Denial of Service. A participant can refuse to sign the valid joint transaction or he/she can spend his/her input before the joint transaction is even completed. Protection from Denial of Service in a decentralized system is hard. The participant who is responsible for the denial of service can be detected in a system where a trusted server is used to implement CoinJoin. But then, using a centralized system requires all parties to trust the system. 
\subsection{Off-Chain Transactions}
As the name suggests, an off-chain transaction happens \say{off} the blockchain. Traditionally, an on-chain transaction modifies the state of the blockchain by getting recorded in the ledger after verification and validation by peer nodes. An off-chain transaction achieves verification and validation with other methods. Although the primary intention behind off-chain transactions was to break the bottleneck limit of Bitcoin transactions, it can account for some amount of privacy as well. This is because, unlike on-chain transactions, not all states of these transactions are recorded on the chain. All the intermediary states of the transactions are stored between the users and only periodic summaries are written to the chain \cite{Medium_Privacy}.  Consequently, tracing and tracking these transactions is much harder. Primarily, these transactions are implemented using Payment Channels. These payment channels between agreed participants allow for multiple Bitcoin transactions to be performed securely, without all of them being validated on the blockchain. On-chain transactions offer a formal verification system of the transactions by placing trust into the Cryptocurrency network, off-chain transactions mutual agreement between both parties or a trusted third party to validate the transactions. The Lightning Network, proposed as an overlay over the Bitcoin network, uses off-chain transactions. 

\subsection{CoinWitness}
Another proposal by Greg Maxwell, which was also known as Pay-to-SNARK allows payment to a user who can produce cryptographic evidence about running a certain deterministic program on a given input argument. This proposal was supposed to be included in the system as a soft-fork without any significant changes. The CoinWitness idea was based on the \cite{ben2013snarks} paper where they constructed a Zero-Knowledge system where anyone can run an arbitrary program inside a specially created environment and publish a quickly verifiable proof that proves - 
\begin{enumerate}
    \item the program was executed without any modification
    \item the program accepted a set of publicly known inputs and non-publicly known inputs and returned true (exited cleanly)
\end{enumerate}
Here, the validator learns nothing apart from the publicly known inputs and the fact that the non-publicly known inputs were \say{accepted}. One of the obvious applications of this was to replace this Zero-Knowledge idea with the Bitcoin script. The size of the cryptographic evidence that was generated was linearly proportional to the input size but constant w.r.t  size and running time of the program. Despite being constant, the size of the proofs was considerably large resulting in large-size proofs being placed in the Bitcoin blockchain network. Consequently, the proposal never became an official BIP but is still on the soft-fork wishlist.

\subsection{CoinSwap}
If two or more parties in the Bitcoin peer network wish to exchange coins, then the traditional way is to go through a trusted third party who is responsible for maintaining fairness in the transaction. CoinSwap allows the parties to achieve the same functionality without trusting the third party. This idea of transacting without trusting the third party is called atomic swaps \cite{han2019optionality}. In Bitcoin, this is achieved through Bitcoin scripts. In this transaction, either the Bitcoin swap/exchange happens for both parties or it does not happen for anyone, thereby maintaining fairness. CoinsSwaps break the transaction graph and prevent inference/analysis attacks. On the blockchain, these transactions appear completely disconnected. However, this system requires a significant amount of interaction between both parties and therefore, is prone to denial of service attacks. CoinSwaps are required to have non-censorship and liveliness requirements from both parties. CoinSwaps can fail and can result in unfairness if these two conditions are not met. Even though this proposal is convincing from a privacy standpoint, it hasn't been deployed yet on the Bitcoin blockchain. 

\subsection{Hierarchical Deterministic Wallets}

Bitcoin and its derivatives have an elegant way of creating new receiving addresses. Hierarchical Deterministic Wallets (HD Wallets) are one of two types of Deterministic Wallets. The other type is Sequential Wallets. A Deterministic Wallet makes use of a seed to derive multiple addresses. The seed in most cases is a human-understandable phrase that can backup and restore a wallet. In Hierarchical Deterministic Wallets, the seed derives a tree of multiple chains of keypairs. On the other hand, Sequential Deterministic Wallets use the seed to derive a chain of keypairs. Suppose a receiving address is to be exchanged with a sender, then it would unintentionally result in sharing all the derived public key addresses if the Sequential Deterministic Wallet model was used. This is not suitable as it reveals unnecessary information. The Hierarchical Deterministic Wallets allow sharing of only the selected keys, thereby protecting the other keys from being revealed. The Bitcoin Improvement Protocol (BIP) 32 provides the specification for Hierarchical Deterministic Wallets and BIP 44 provided improved implementation of the same.\\

\begin{figure}[hbt!]
\centering
\includegraphics[width=2.5in]{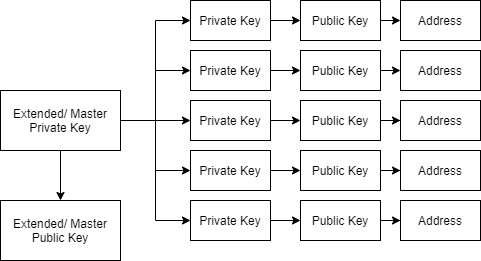}
\caption{Hierarchical Deterministic Wallets}
\label{hdwallets}
\end{figure}

The recipient has an extended private key which is the master private key from which all the other private keys for the recipient are derived. A chain code $c$ is an added 256 bits of entropy that also contributes towards generating the private keys. The private keys generated from the extended private key, which is represented as $(k,c)$ would be used to generate the corresponding public keys which are then used to generate the addresses. So, the user's account has multiple public-private key pairs that are derived from the same extended private key which is generated from the seed phrase. The seed phrase, extended or master private key, and all the derived private keys must be kept secret. A tree-like hierarchical structure of keys is formed where the extended private key is the root. This is shown in \ref{hdwallets}. The generation of keys is deterministic. An extended or master public key is generated from the extended private key. It is represented as $(K,c)$. The extended public key cannot generate the private keys in the tree, but it can generate the public keys. The extended public key is not kept secret. \\

The Hierarchical Deterministic Wallet scheme is used as follows:

\begin{enumerate}
    \item The extended public key $(K,c)$ is shared between the sender and recipient.
    \item The sender selects a 32 bits integer $i$.
    \item The sender uses $(K,c,i)$ to generate a public key $K_i$. 
    \item The sender includes $K_i$ as the recipient address in the transaction.
    \item The recipient can check if the transaction was successfully executed by checking the balance at $K_i$. $i$ should be a pre-determined value between the sender and the recipient otherwise the recipient would have to search for the transaction for all possible values of $i$, which is $2^{32}$.
    \item For future transactions to the same recipient, the sender chooses another value for $i$ that is increased sequentially for generating a new one-time public key corresponding to that $i$.
\end{enumerate}
The public key $K_i$ acts as a one-time public key which is unique for every transaction that a sender sends. If one sender sends multiple transactions to the same recipient, it would use different public keys. From the blockchain, given all the transaction data, inferring the existence of multiple transactions between a sender and a receiver is not possible due to the use of the one-time public keys. \\

However, the Hierarchical Deterministic Wallet fails to provide privacy in the following ways:

\begin{enumerate}
    \item Sharing the extended public key $(K,c)$ to a user allows the users to generate the entire list of public keys. The user may simply use this list and iterate through each public key to view the transactions that include them. The transaction patterns would be clearly visible.
    \item BIP 32 suffers from a vulnerability that makes it possible to generate the extended private key given the extended public key and any of the child private keys\cite{Buterin}.
\end{enumerate}

The deterministic nature of these wallets is problematic as they can be leveraged to conduct various sorts of attacks. Moreover, the extended public keys do not protect the privacy of the entire tree of public keys. They must be handled with more diligence than a child public key.\\

A Hierarchical Deterministic Wallet can be used when a large organization wants to allow its various departments to use cryptocurrencies to receive payments. The master private key is kept secure with the administrator and each branch is given a private key. The branches would then generate the public key corresponding to their private keys. The extended or master public key can be shared with all the departments. This model is suitable as all the departments can operate without having to interfere with each other as all the receipts are collected separately. However, there is a security flaw in this model. The vulnerability, which is also publicly specified in the documentation of BIP 32, allows any department having a private key to generate the master private key, given the master public key. This does not provide the separation that we wanted and the organization is not protected against insider attacks. In this situation, usage of 2-out-of-3 multi-sig transactions may be better. However, no research has been done to formally prove it.

%\subsection{ZeroCoin (Proposed but not accepted)}

\section{Privacy-Preserving Techniques Used in Other Cryptocurrencies}
\subsection{Ring Signatures}
A user who wants to generate a ring signature is assigned a set. The set consists of multiple users whom all have public-private key pairs. The user generates the ring signature on a message by knowing the public keys belonging to the rest of the members of the set. Coordination among the members of the set is not required, nor is there any need for a centralized authority, unlike CoinJoin. The other members do not need to be aware of their involvement in the signature. None of the members can forge each other's signature. The ring signature is irrevocable. The verification of the ring signature can be done with the knowledge of the message, the ring signature, and all the public keys owned by the members of the set. The signer's identity remains hidden behind this set\cite{log_size}. The goal of using ring signatures in Cryptocurrencies is to protect the privacy of a sender of a transaction by making it computationally infeasible to determine the sender's address given the signature\cite{bender2006ring}.\\

\begin{figure}[hbt!]
\centering
\includegraphics[width=2.5in]{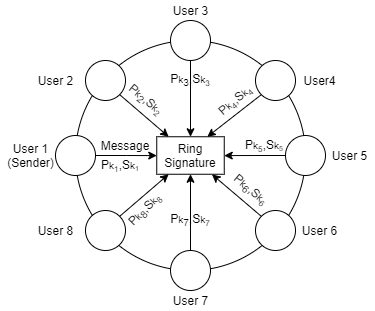}
\caption{Ring Signatures}
\label{ringsignature}
\end{figure}

Ring signatures were invented by Ron Rivest, Adi Shamir, and Yael Tauman Kalai with the aim of anonymizing confidential information, protecting the identity of a signer and only revealing it to the intended recipient and utilizing it for secure multiparty computations\cite{Leak_Secret}. Ring signatures have found their use in privacy-oriented Cryptocurrencies. It is used in ShadowCash, which originally implemented it incorrectly making it possible to de-anonymize the identities of its users. CryptoNote implemented a modified version of ring signatures, called traceable ring signatures, that prevented double-spending attacks by prohibiting a sender from signing two different ring signatures using the same public key without having gained attention from the blockchain. The validating nodes would reject such a transaction and overall, the consensus protocol would prevent double-spending. RingCoin also made use of a different type of ring signatures, called linkable ring signatures, that are a slight modification of
the Linkable Spontaneous Anonymous Group signatures.\\

Monero uses ring signatures and enforces all the members in the set to hold the same amount of coins which is equal to the number of coins to be consumed in the transaction. The members other than the real sender pull up this amount arbitrarily from the blockchain, making the inputs of the transaction untraceable. Monero makes use of key images that are associated with every ring signature to avoid double-spending. Since the signature does not reveal the sender's identity and the balance of all the members looks the same, the identity of the sender of a transaction is hidden. A diagrammatic representation of ring signatures can be shown in fig. \ref{ringsignature}\\

Ring signatures have a requirement that all the members on the ring must have the same outputs. The value of this output is not concealed. Even though ring signatures are successful in hiding the identity of the sender or the signer of the transaction, they fail in hiding the amount transferred in the transaction. This piece of information can be used as a clue to deduct inferences like transaction patterns. For instance, if a target, who is the sender whose identity an attacker wishes to uncover, always pays a recipient who does not make use of stealth addresses and whose public address is known to the attacker, he/she would be able to find this transaction by searching for transactions that transfer the specified amount to the recipient. The time when the transaction is generated can also aid in narrowing down the target transaction from several potential transactions. Even though the sender's address stays concealed, due to the use of ring signatures, the transactions that the sender makes for the specified amount can still be found. The attacker can make many inferences based upon this information, such as finding the frequency of such transactions and can potentially determine the address of the sender. More importantly, a sender may want to keep the transaction amount private. Ring Confidential Transactions help achieve this goal. \\

If the transaction amount is a less common one, then the set that manages to pull up this amount may be small. A smaller set would fail to guarantee anonymity to the sender. Furthermore, ring signatures do not prevent the formation of dust transactions.

%RingCTs does not allow a user to send a dust amount, therefore, preventing the formation of dust transactions, which ring signatures could not control. Ring signatures fail to guarantee signer anonymity when the user set is small but since RingCTs are used as a replacement for ring signatures, this problem does not occur. 

\subsection{Stealth Addresses}
As the transaction data on the blockchain is public it is easily feasible to determine the recipient wallet address where the transaction outputs are spent. Although pseudonymity is provided which means the blockchain hides the real-world identity of the user that owns the public address, observing transaction patterns can lead to linking the wallet address to the real-world identity of the user. Furthermore, if the user's wallet address and the associated real-world identity is already made public, for instance, if the user is a vendor that accepts payment on its public address, then all the payments that the vendor receives would be observed on the blockchain. The details about consumer orders such as frequency, peak ordering times, and other consumer analytics would be publicly available. The vendor may not want this data to be public. For instance, a competing vendor can use this information to study the vendor's business model and can strategize against this vendor's business. The vendor's trust in the blockchain can be lost if his business is being affected. But, even if the vendor does not suffer any losses, the nature of the information can be labeled as private according to the vendor. \\

\begin{figure}[hbt!]
\centering
\includegraphics[width=2.5in]{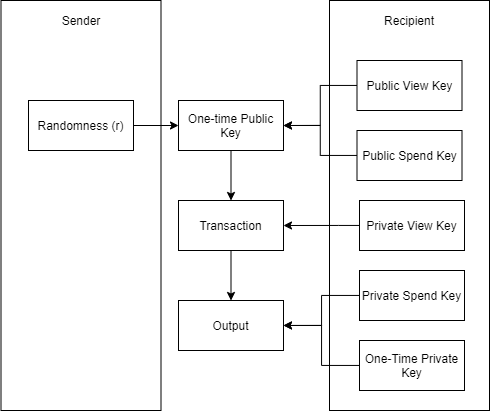}
\caption{Stealth addresses in Monero}
\label{stealthaddresses}
\end{figure}

For protecting a recipient's privacy, the use of a stealth address, which is a privacy-preserving cryptographic technique, is encouraged. A stealth address is a key pair used by a recipient to prevent disclosing its public wallet address. The stealth address scheme guarantees unlinkability of the stealth address and the public wallet address of the recipient\cite{Blockchain_stealth}. A diagrammatic representation of this can be shown in \ref{stealthaddresses}.\\

The stealth address scheme that is followed in Monero is as follows:
\begin{enumerate}
    \item The recipient has a public view key and a public spend key. These keys form the stealth address.
    \item The sender makes use of these keys along with some randomness, which makes it unpredictable and unlinkable with the recipient's key pair, to generate a one-time public key, which would be used as the recipient address in the transaction. The one-time public key can be said to be a temporary destination address.
    \item The one-time public key can be viewed by every other user in the blockchain. But, no one can learn the sender and recipient that share it.
    \item The recipient generates a private view key for scanning the blockchain to find the transaction that was sent by this sender and for retrieving the output of the transaction into their wallet.
    \item The recipient would also compute a one-time private key that corresponds to the one-time public key that allows them to spend the output using their private spend key. Since no one but the recipient can generate the one-time private key, no one can spend the coins that were sent to the corresponding one-time public key.
\end{enumerate}

The one-time public key is unique for every transaction. Therefore, even if multiple transactions are made from the sender to the same recipient, it is not possible to prove the existence of such transactions with the use of the stealth address scheme. The recipient's wallet address is never specified in the transaction and is hidden with the help of the one-time public key. One cannot infer the recipient's wallet address from the one-time public key even though the one-time public key is easily accessible to anyone. Moreover, the outputs of the transaction can only be consumed by the intended recipient without leaking any additional information. This unlinkability between the one-time public key and the recipient's wallet provides privacy to the recipient.\\

Only one stealth address can be associated with the recipient's public address. If many senders want to transfer coins to the recipient, the stealth address is shared with each of them. Using this stealth address as the identifier, the senders can find that their intended recipient is the same. This is an attack on the privacy of the recipient's stealth address. A trivial and unappealing solution would be one where the recipient generates a new wallet address for every transaction or each unique sender. This is unappealing because the use of stealth addresses in itself acts as a way to eliminate generating new addresses every time due to the computation of a unique one-time public key for each transaction. Generating new addresses each time would although provide privacy, but managing these addresses can be difficult. However, if a new public address is generated for each sender, where the number of senders is a small finite number can be an appealing solution because managing the public keys would not be so difficult.  

\subsection{Rings Confidential Transactions}
Ring Confidential Transactions (RingCTs) is a privacy-preserving technique that hides the amount of coins transferred from the sender to the receiver in a transaction. Monero makes use of Multilayered Linkable Spontaneous Anonymous Group Signatures (MLSAG) for combining ring signatures and Confidential Transactions for preventing double-spending and protecting the anonymity of the sender and receiver and hiding the transaction amount efficiently\cite{Ring_Confidential}. MLSAGs are almost similar to Linkable Spontaneous Anonymous Group Signatures (LSAG) with the difference being in the use of a key vector, which is a collection of public-private key pairs, in MLSAGs. MLSAGs provide the following proofs:

\begin{enumerate}
    \item In a ring containing $n$ members, a signer can create a valid signature only if the $m$ private keys belonging to the members' key vectors are known to the signer. MLSAGs are unforgeable under the discrete logarithm assumption.
    \item Creating two different signatures $\sigma_1,\sigma_2$, using two different key vectors $\bar{y_1},\bar{y_2}$ that both share a public key would result in the two rings being linked by key image. This is the proof of linkability in MLSAGs and is used to detect double-spending attacks.
    \item MLSAGs provide signer ambiguity by making it computationally difficult to determine the signer of a verifiable signature given a set of key vectors under the Decisional Diffie-Helman assumption. 
\end{enumerate}

The Pedersen Commitment scheme is used to create commitments for the inputs and outputs. A ring signature consisting of the input commitments added with the corresponding public keys and subtracting with the sum of the output commitments is created. The input commitments are added with the corresponding public keys to create a pair of the commitment and the public key that allows them to be spent together. The commitment can be verified to prove that the sum of inputs equals the sum of outputs, without revealing the actual input and output values. The ring signature guarantees sender anonymity and is associated with a key image to detect linking of two commitments, which is an indication of a double-spending attack.\\ 

Aggregate Schnorr Non-linkable (ASNL) ring signatures are used to provide range proofs for output values. Given the output commitments, it can be proved that the output values are positive and lie in a specified range $[0,2^n]$, where n is any number. ASNL ring signatures are unforgeable under the discrete logarithm assumption. Since Ring Confidential Transactions are used for concealing the transaction amount, user transaction analysis through dust attacks would be unsuccessful. The creation of a dust attack can be prevented if the range proof algorithm can prove that a dust amount was not sent in the transaction.

\subsection{Z-Addresses}
\begin{figure}[hbt!]
\centering
\includegraphics[width=2.5in]{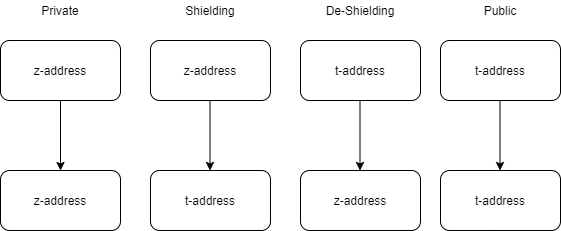}
\caption{ZCash Transaction Types}
\label{zcash}
\end{figure}

Zcash is a cryptocurrency that utilizes zero-knowledge proofs in building privacy-preserving techniques. For using this cryptocurrency, Zcash provides users with two addresses: z-addresses and t-addresses. Z-addresses are private addresses whereas t-addresses are public addresses. Z-addresses are also called shielded addresses and t-addresses are called transparent addresses. Whenever a user wants to keep their activity private, z-addresses would guarantee it. Any activity that the user wants to keep public can be done using the t-address. Interoperability between the z-addresses and t-addresses is assured. Figure \ref{zcash} shows the possible types of transactions in Z-Cash. Since the z-addresses and t-addresses are interoperable, four different types are possible. The privacy of the transaction depends on the addresses used by the sending and receiving party. \\

Transactions involving z-addresses or shielded addresses make use of zero-knowledge proofs to conceal the sender and receiver's addresses in a transaction, the amount transferred, and the memo fields. The transaction appears in encrypted form on the blockchain and can be verified by validating nodes. 

% \subsection{CryptoNote}
% \subsection{Tor Network}
% \subsection{Bullet Proofs}
\subsection{zk-SNARKs}
zk-SNARKs (Zero-Knowledge Succinct Non-Interactive Argument of Knowledge), when applied to cryptocurrencies, act as a privacy-preserving technique. It involves a prover and verifier, where the prover using zk-SNARKs, proves to the verifier that they possess a certain piece of information without actually revealing that information to the verifier. Succinct refers to the small proof size, which is about only a hundred bytes irrespective of the size of the program, and small verification time, which is often in milliseconds. Non-interactive refers to the communication method of the proof which is limited to a single message that is sent from the prover to the verifier. Any further interaction is not required.\\

Zcash has implemented zk-SNARKs to provide privacy to transactions that involve shielded addresses. Zcash uses an initial setup phase to establish a common reference string which is a set of parameters that are known to both the prover and the verifier for constructing and verifying zk-SNARK proofs. These parameters are called public parameters of the system and are readily made available on the blockchain. Proving and verifying keys are shared with all nodes using the public parameters of the system. The sender of a shielded transaction uses a proving key to construct a proof which is verified by a validating node using a verifying key\cite{zksnarks}. The zk-SNARK proofs allow the validating nodes to validate a shielded transaction without revealing private information in the transaction, such as, the sender and receiver addresses, transaction amount, etc. \\

The zk-SNARK proof is constructed by the sender of the transaction, who uses a shielded address, to prove that:
\begin{enumerate}
    \item The sum of inputs to the transaction equals the sum of output generated by the transaction, for every shielded transfer.
    \item they possess the private spending keys for the inputs, which means that they have the authority to spend them.
    \item The private spending keys are cryptographically linked to the transaction signature, which means that the transaction cannot be modified by someone who does not possess the private spending keys.
\end{enumerate}

Each output that is generated in a shielded transaction is called a commitment. The commitments are made available to all the nodes in the blockchain. The commitment is stored as a hash to conceal the information about the transaction outputs. It is of the form: Hash(recipient address, amount, $\rho$, r), where recipient address is the address to which the output is destined, \say{amount} is the amount that is transferred to the recipient, rho is a unique value used to derive a nullifier, and r is a random nonce. A nullifier is created and made available to all the nodes in the blockchain. It helps in proving that the sender, that created the commitment, has the authority to spend the amount specified in the commitment. The sender generates a nullifier, which is of the form: Hash(spending key, rho), where spending key is a private key that the sender possesses that gives them the ability to spend the amount, and rho is the unique value associated with the unspent commitment. The commitment is revealed when the nullifier is presented. A revealed commitment means that the sender has spent the amount\cite{Terrence}.\\

To prevent double-spending, a zero-knowledge proof is constructed which is verified by the validating nodes to ensure that a revealed commitment exists for every input to the transaction, the commitments and nullifiers for the outputs of the transactions are generated correctly and the nullifiers of all the commitments are unique. \\

Based on public parameters of the system that are created in the initial setup phase, zk-SNARK proofs are constructed and validated. If enough randomness is not introduced in their generation, an adversary can forge proofs and create counterfeit coins. However, the privacy of users, who use shielded transactions, is still protected even if the initial setup phase fails or the random numbers are revealed. To prevent this from happening, Zcash uses multi-party computation ceremonies to generate these parameters. All parties contribute towards computing the parameters, and the ceremony protects against parties that may be corrupted. For the ceremony to fail in providing security to the final public parameters, all parties involved would have had to be dishonest or corrupted\cite{paramgen}.

\begin{table*}[hbt!]
\begin{center}
\begin{tabular}{|p{0.13\linewidth}|p{0.10\linewidth}|p{0.10\linewidth}|} 
\hline
Technique & 
User Privacy & 
Anonymity Set \\ \hline
Untrusted Mixer   &
Low & 
Large \\ \hline
MixCoin &
Low & 
Large \\\hline
BlindCoin &
High & 
Large \\\hline
CoinJoin &
Low & 
Small \\\hline
CoinSwap &
Low & 
Large \\ \hline 
\end{tabular}
\end{center}
\caption{Comparing various Privacy-Preserving Techniques in Bitcoin}
\label{table1}

\end{table*}

\begin{table*}[hbt!]
\begin{center}
\begin{tabular}{|p{0.1\linewidth}|p{0.07\linewidth}|p{0.07\linewidth}|p{0.07\linewidth}|}
\hline
Technique & 
Sender Privacy & 
Recipient Privacy & 
Amount Privacy \\ \hline
Ring Signatures &
Limited & 
No & 
No \\ \hline
Stealth Addresses &
No & 
Strong & 
No \\\hline
Pedersen Commitments &
No & 
No & 
Strong \\\hline
zk-SNARKs &
Strong & 
No & 
Strong \\\hline
Ring Confidential Transactions &
Limited & 
Strong & 
Strong \\ \hline 
\end{tabular}
\end{center}
\caption{Comparing various Privacy-Preserving Techniques in Cryptocurrencies other than Bitcoin}
\label{table2}

\end{table*}

\begin{table*}[hbt!]
\begin{tabular}{|p{0.18\linewidth}|p{0.07\linewidth}|p{0.07\linewidth}|p{0.07\linewidth}|p{0.07\linewidth}|p{0.07\linewidth}|p{0.07\linewidth}|p{0.2\linewidth}|}
\hline
 &
  Bitcoin &
  Ethereum &
  Monero &
  Dash &
  Verge &
  ZCash &
  Bitcoin + Lightning Network \\ \hline
Analysis of Ledger &
  Possible &
  Possible &
  Partially Possible &
  Possible &
  Possible &
  Possible if TX is unshielded &
  For opening/closing states \\ \hline
Sender Address of Transaction &
  Public &
  Public &
  Private &
  Public &
  Public &
  Private &
  Private outside channel, public within channel \\ \hline
Recipient Address of Transaction &
  Public &
  Public &
  Public but unlinkable &
  Public (can be made unlinkable) &
  Public but unlinkable &
  Private &
  Private outside channel, public within channel \\ \hline
Transaction Amount &
  Public &
  Public &
  Private &
  Public &
  Public &
  Private &
  Opening/closing states are public but inner states are private \\ \hline
List of Addresses &
  Public &
  Public &
  Private &
  Public &
  Public &
  Private &
  Public \\ \hline
Balances / Smart Contract Code &
  Public &
  Public &
  Private &
  Public &
  Public &
  Private &
  Opening/closing states are public but inner states are private \\ \hline
Relationship Between Sender and Receiver &
  Public &
  Public &
  Private &
  Public &
  Public &
  Private &
  Private outside channel, public within channel \\ \hline 
\end{tabular}\\
\caption{Comparison of Privacy in Various Cryptocurrencies}
\label{table3}
\end{table*}

%https://www.tablesgenerator.com

\section{Comparing privacy in various cryptocurrencies}
Comparing privacy provided by different cryptocurrencies precisely is not an easy task\cite{genkin2018privacy}. Comparison and analysis of privacy in various cryptocurrencies can be achieved by separately analyzing the privacy of various aspects involved in the cryptocurrency, such as:
\begin{enumerate}
    \item the identity of the users (sender and the recipient) involved in a transaction,
    \item the transaction data,
    \item the entire available blockchain ledger state resulting from the transactions\cite{Medium_Privacy}.\\
\end{enumerate}

By answering questions like:
\begin{enumerate}
    \item Is the sender's address private?
    \item Is the receiver's address public?
    \item Can the amount transferred in the transaction be found?
    \item Would a small anonymity set guarantee privacy?
    % \item Does a vulnerability exist in the implementation of Hierarchical Deterministic Wallets that can reveal the public addresses?
\end{enumerate}
we can determine the guarantee of privacy precisely. The ability of privacy-preserving techniques in providing privacy at times may depend on the underlying cryptographic schemes, such as in stealth addresses, hierarchical deterministic wallets. It may also depend on the anonymity set or the centralized or decentralized nature of the protocol. A large anonymity set typically is better for privacy. Decentralized protocols eliminate the need for a trusted third party, which can be a single point of failure. However, the anonymity set that decentralized protocols can provide depends on the number of active participation and the coordination between them. Typically, the anonymity set is small. Therefore, decentralized protocols provide weaker privacy compared to centralized protocols that can manage to gather a large anonymity set at any point in time. In Table \ref{table1}, we compare the privacy provided by privacy-preserving techniques in Bitcoin, along with their typical anonymity set size.\\

Users should be made aware of the conditions that need to be met for their transaction to be called private. They must not be under the false assumption of privacy. For instance, CoinJoin can provide privacy only when the anonymity set is large. Generally, only 2–4 participants on average are observed to participate in a CoinJoin transaction. As a result, researchers could de-anonymize 67\% of CoinJoin transactions\cite{Chen_2019}. In Zcash, users should remember to use z-addresses or shielded addresses for transactions that they want to keep private. It is important to understand what technique provides privacy for a certain aspect. For instance, stealth addresses cannot be used to conceal a sender's identity, and ring signatures cannot hide the receiver's identity. This should, however, not be considered as a flaw in stealth addresses. They are designed for a specific goal, and attainment of only that goal should be verified. In Table \ref{table2} we provide a comparison of the privacy-preserving techniques in cryptocurrencies other than Bitcoin in terms of the privacy guarantees they provide. Finally, we compare privacy provided by various cryptocurrencies in terms of certain parameters in Table \ref{table3}. \\  

While attempting to de-anonymize or conduct inference attacks on transactions, attackers may possess additional information that they might have obtained from other sources than the blockchain. The use of this additional information coupled with de-anonymization and inference attacks may result in attacks on the privacy of the transactions even if the blockchain and the user involved in the transaction makes use of privacy-preserving techniques. The theoretical guarantee of privacy may differ from the practical guarantee of privacy. For instance, researchers conducted an attack on Monero that leveraged the fact that the output does not remain unspent for an infinite time. Its probability of being spent increases with time. Their attack strategy is defined in the following manner: Given a set of input keys used to create a ring signature, the real key being spent is the one with the highest block height, where it previously appeared as an output. Their attack had a true positive
rate of 98.1\% which shows that very often the most
recent output is the real one being spent\cite{kumar2017traceability}.

\section{Legality of privacy-preserving cryptocurrencies}
Many cryptocurrencies that offer complete anonymity or transaction anonymity have not been widely accepted due to their primary use-cases being money laundering and other illicit activities. Cryptocurrencies like Monero (XMR) are not listed on crypto-exchanges due to their hard-to-track privacy features. Often, fewer people tend to get involved in the network if the means of converting the cryptocurrency into a fiat currency are prohibited. These cryptocurrencies are constantly monitored by government agencies to track trafficking and other illegal activities. Even though there are ways to exchange these currencies to Bitcoin or other popular cryptocurrencies, they immediately get tainted as inter-currency transactions are managed mostly by crypto-exchanges. There's always been a trade-off between acceptability and privacy when it comes to cryptocurrencies. The optimal way out of this would be to adhere to a moderately private tier of privacy and yet maintain acceptability. Soft forks in Bitcoins that have attempted to implement privacy measures have managed to maintain the legality and wide acceptability of the cryptocurrency in many countries. 

\section{Conclusion}
In this paper, we distinctly identified the tiers of privacy and compared prominently known Cryptocurrencies based on their privacy offerings. We also studied various privacy-preserving measures and algorithms proposed to satisfy the two primary properties of untraceability and unlinkability. Several privacy problems and various inference analysis techniques on Bitcoin were surveyed. As countermeasures, several schemes were incorporated into the Bitcoin network via overlaying implementations or soft-forks that tackled many of the privacy problems that were found in the Bitcoin network. We also studied privacy-preserving techniques offered by other Cryptocurrencies and compared them based on the algorithms and techniques used to conceal information and prevent analysis attacks.

\bibliographystyle{unsrt}
\bibliography{citations.bib}

% %   URLS
% % https://eprint.iacr.org/2015/1098.pdf
% https://www.iacr.org/archive/asiacrypt2001/22480554.pdf
% % https://eprint.iacr.org/2019/196.pdf
% % https://eprint.iacr.org/2019/135.pdf
% % https://eprint.iacr.org/2005/304.pdf
% % https://www.scitepress.org/Papers/2017/62700/62700.pdf
% % https://eprint.iacr.org/2020/548.pdf
% % https://en.bitcoin.it/wiki/BIP_0032
% % https://bitcoinmagazine.com/technical/deterministic-wallets-advantages-flaw-1385450276#:~:text=Deterministic%20Wallets%2C%20Their%20Advantages%20and%20their%20Understated%20Flaws,-Vitalik%20Buterin%20Nov&text=Unlike%20old%2Dstyle%20Bitcoin%20wallets,algorithm%20from%20a%20single%20seed.

% https://www.gemini.com/cryptopedia/crypto-dusting-attack-bitcoin
% https://coinmarketcap.com/alexandria/glossary/dust-transactions
% https://coinrivet.com/dustaid/
% https://dl.acm.org/doi/pdf/10.1145/3132696
% https://arxiv.org/pdf/1706.00916.pdf

% % that's all folks
\end{document}